\documentclass[lettersize,journal]{IEEEtran}

\usepackage{comment}

\begin{comment}
\documentclass{IEEEtran}
\usepackage{comment}
\usepackage{silence}
\usepackage{cite}
\usepackage{amsmath,amssymb,amsfonts}
\usepackage{algorithmic}
\usepackage{graphicx,color}
\usepackage{textcomp}
\usepackage{upgreek} 
\usepackage{siunitx} 
\usepackage{graphicx}
\usepackage{cite}
\renewcommand\IEEEkeywordsname{Keywords}

\def\BibTeX{{\rm B\kern-.05em{\sc i\kern-.025em b}\kern-.08em
    T\kern-.1667em\lower.7ex\hbox{E}\kern-.125emX}}
\AtBeginDocument{\definecolor{ojcolor}{cmyk}{0,1,0.63,0.12}}
\def\OJlogo{\vspace{-10pt}\includegraphics[height=22pt]{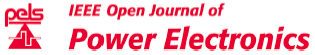}}

\title{\Large \bf Estimation of Semiconductor Power Losses Through Automatic Thermal Modeling}
\author{José Miguel Sanz-Alcaine$^1$, 
Eduardo Sebastián$^2$, 
Francisco José Perez-Cebolla$^1$,
\\ Asier Arruti Romero$^3$,
Carlos Bernal-Ruiz$^1$,
Iosu Aizpuru$^3$
\thanks{This work has been supported by the Spanish project CDTI - MIG-20201042, a DGA Ph.D. grant, and a Spanish Ph.D. grant FPU19-05700.}%
\thanks{$^1$GEPM research group, Aragón Institute of Engineering Research (I3A), University of Zaragoza, Zaragoza, Spain}%
\thanks{$^2$RoPeRT research group, Aragón Institute of Engineering Research (I3A), University of Zaragoza, Zaragoza, Spain}
\thanks{$^3$Department of Electronics and Computing, Faculty
of Engineering, Mondragon Unibertsitatea, Arrasate/Mondragon 20500, Spain}
\thanks{Corresponding author: José Miguel Sanz-Alcaine (jm\_sanz@unizar.es)}
}

\begin{document}
\maketitle





\begin{abstract}
The optimal design of power converters requires accurate knowledge of the dissipation elements of its system to achieve the desired performance and security requirements. Calorimetric methods have surpassed classical electrical methods for the estimation of semiconductor power losses but have mechanical limitations and resort to analytical electrothermal equivalent circuits for this task. These electrothermal models are highly dependent on the topology and technology used on the power converter leading to either simplifications that underestimate the thermal effects or intractable sets of differential equations. To overcome these issues, we propose a novel data-driven identification method to characterize the thermal dynamics of power converters allowing the designer to obtain semiconductor total power losses only by means of temperature measurements without the need of a calorimeter. Given a set of power vs.temperature profiles, our solution identifies the linear model that best fits the data. The solution is based on an optimization problem that allows not only accurate identification but also coding of the desired modeling requirements, such as dynamics' invertibility to allow the estimation of power losses from the temperature profiles. The proposed methodology can be applied to any power converter topology. Furthermore, by obtaining a linear model, standard control theory techniques can be exploited to analyze and control the thermal dynamics. Real experiments validate the generality and accuracy of the proposal.
\end{abstract}


\maketitle

\section{Introduction}\label{sec:intro}
\input{00_Introduction}

\section{Problem Formulation}\label{sec:prosta}
\input{01_Problem}

\section{Proposed Solution}\label{sec:modeling}
\input{02_Modeling}

\section{Characterization Methodology}\label{sec:CharMethod}
\input{03_CharacterizationMethodology}

\section{Automatic Thermal Modeling Results}\label{sec:setup}
\input{04_AutomaticResults}

\section{Conclusion}\label{sec:conclusions}
\input{05_Conclusions}

\bibliographystyle{IEEEtran}
\bibliography{refs.bib}

\end{document}